\documentclass{rspublic}  
\usepackage{graphicx}

\usepackage{amssymb}
\usepackage{amsmath}
\usepackage{times}
\usepackage{color}

\usepackage[numbers,sort&compress]{natbib}

\begin{document} 
\title[Multiple mutations]{A deterministic model for the occurrence and dynamics of multiple mutations in hierarchically organized tissues} \author[BW, DD, AT]{Benjamin Werner$^{1}$, David Dingli$^{2}$, Arne Traulsen$^{1}$} \affiliation{$^{1}$Evolutionary Theory Group, Max-Planck-Institute for Evolutionary Biology, Pl\"on, Germany, $^{2}$Division of Hematology, Mayo Clinic, College of Medicine, Rochester, MN, USA} \label{firstpage} 

\maketitle

\begin{abstract}{
Cancer, clonal extinction \& diversity, mathematical modeling}
Cancers are rarely caused by single mutations, 
but often develop based on the combined effects of multiple mutations. 
For most cells, the number of possible cell divisions is limited due to various  biological
constrains, as for example progressive telomere shortening, cell senescence cascades or a hierarchically organized tissue structure.
Thus, the risk of accumulating cells carrying multiple mutations is low.
Nonetheless, many diseases are based on the accumulation of such multiple mutations.
We model a general, hierarchically organized tissue by a multi compartment approach, allowing any number of mutations within a cell. 
We derive closed solutions for the deterministic clonal dynamics 
and the reproductive capacity of single clones.
Our results hold for the average dynamics in a hierarchical tissue characterized by an arbitrary combination of proliferation parameters.
We show that hierarchically organized tissues strongly suppress cells carrying multiple mutations and derive closed solutions for the expected size and diversity of clonal populations founded by a single mutant within the hierarchy. 
We discuss the example of acute lymphoblastic childhood leukemia in detail and find a good agreement of our predicted results and recently observed clonal diversities in patients.
This result can contribute to the explanation of very diverse mutation profiles observed by whole genome sequencing of many different cancers. 
\end{abstract}

\section{Introduction}

The lifespan of most cells in biological organisms is limited and usually the life expectancy of the organism exceeds this time by orders of magnitudes \citep{hayflick:ECR:1961,raff:Nature:1992}. As cells are continuously lost, mechanisms to replenish the cell pool evolved in organisms, enabling sustained cell production during life time \citep{pardee:Science:1989}. Often this is realized by hierarchically organized tissue structures. At the root of the hierarchy are few tissue specific stem cells, combining two properties, self renewal and differentiation potential \citep{mcculloch:NM:2005}. 
During cell proliferation, cells differentiate and become increasingly specialized to perform specific functions within the hierarchy. After some differentiation steps the complete spectrum of functional cells can be obtained \citep{michor:PRSB:2003,dingli:PlosOne:2007,nowak:book:2006,michor:Nature:2005,wodarz:book:2005,werner:PlosCB:2011}. A prominent example is the hematopoietic system \citep{loeffler:CTK:1980,michor:Nature:2005,dingli:PlosOne:2007,nowak:book:2006,wodarz:book:2005,werner:PlosCB:2011}, but other tissues as for example skin \citep{fuchs:JCB:2008,tumbar:Science:2004} or colon \citep{potten:CP:2009} are also hierarchically organized.  

A large number of cell divisions are indispensable to life, however, they are unavoidably  accompanied by mutations. Typically, these cells are washed out of the hierarchy and thus, especially if they arise in relatively differentiated cells, the associated mutations 
are lost in the long run \citep{araten:PNAS:1999,werner:PlosCB:2011}. 
But cells with multiple mutational hits might persist for a long time, 
causing the risk of the accumulation of additional mutations during cell proliferation, which can ultimately lead to cancer. 
Although some cancers seem to be caused by single mutation hits, for example the BCR-ABL translocation occurring in stem cells in chronic myeloid leukemia \citep{daley:Science:1990} or the PML-RARA translocation occurring in more differentiated cells in acute promyelocytic leukemia \citep{guibal:Blood:2009}, they are rare. The majority of cancers are triggered by at least a handful of mutations \citep{stratton:Nature:2009,beerenwinkel:PlosCB:2007,gerstung:MPS:2010}. The recent progress in genome sequencing techniques allowed in some cases the classification of cancer initiating mutations, in other cases the underlying mutations remain unknown \citep{welch:Cell:2012}. However, many of these studies reveal a very diverse mutation landscape, indicating the existence of several cancer initiating driver mutations and additional alterations that have a small or even no impact on cancer development, so called passenger mutations \citep{hanahan:Cell:2000,vogelstein:NM:2004,sjoblom:Science:2006,jones:PNAS:2008,pleasance:Nature:2010,Shah:nature:2012,ding:Nature:2012,walterNEJM:2012}. The precise impact of passenger mutations on cancer progression is still under discussion, but the 
typical assumption is that they are
neutral and do not affect the proliferation properties of cells \citep{haber:Nature:2007,bozic:PNAS:2010}. In particular, this holds for synonymous mutations that do not have any consequences on protein structure or function \citep{greenman:Nature:2007}.

In mathematical and computational approaches, 
compartment models are frequently used to describe cell dynamics in hierarchically organized tissue structures. Many of these studies investigate effects of stem cell mutations and related clinical implications, see for example \citep{michor:Nature:2005,brenes:PNAS:2011,roeder:NM:2006,dingli:PlosOne:2007,marciniak-czochra:SCD:2009,glauche::AgingCell:2011}. Also stochastic effects of tissue homeostasis are analyzed \citep{lenaerts:HAE:2010,Lenaerts:CCY:2011,traulsen:JRSI:2013}, highlighting, that cancer driving mutations can in principle disappear by chance (stochastic extinction). The interplay of stem cell and progenitor cell mutations and their impact on cancer initiation are discussed \citep{komarova:CCY:2004} and game theoretical approaches allow to model evolutionary aspects of tissue homeostasis and inter cell competition \citep{basanta:Phys-Biol:2011,gerlee:Prog-Biophys-Mol-Biol:2011}. Often, these studies investigate the effects of cells carrying one or very few specific mutations and assume either constant population size or only minimal hierarchies. 

Here, we focus on the presence of cells carrying multiple mutations within a hierarchically organized tissue.
We show mathematically, that the hierarchical organization strongly suppresses cells carrying multiple mutations and thus reduces the risk of cancer initiation. Closed solutions for the total cell population that arises from a single (mutant) cell are derived and from this the expected diversity of the mutation landscape and the clonal size can be described. This enables a better understanding of the expected diversity in mutation landscapes that are observed in both healthy and cancerous tissues.

\subsection{Mathematical model}
 
\begin{figure}
\center
    \includegraphics[width=0.6\textwidth]{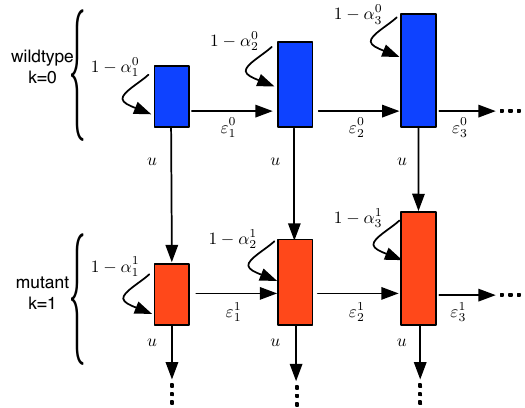}
\caption{Schematic representation of the compartment structure of multiple mutations and corresponding transition rates. The top compartments (blue) contain cells carrying no mutation. The bottom compartments (red) contain cells carrying one mutation. Compartments to the right represent more specialized cell stages and arrows transition probabilities, where $\varepsilon$ denotes the differentiation probability, $u$ the mutation rate of cells and $\alpha_{i}^{k}=\varepsilon_{i}^{k}+u$. Initially no mutated cells are present in the hierarchy. We then ask, how many cells are acquired from the founder compartment (top left) and investigate how many cells with $k$ mutations are on average expected at any stage of the hierarchy.}
\label{Network}
\end{figure}

 The hierarchical tissue organization is typically modeled by a multi compartment approach
 \citep{dingli:PlosOne:2007, werner:PlosCB:2011}. 
 Each compartment represents a certain differentiation stage of cells. 
 At the root of the hierarchy are stem cells ensuring a continuous influx of cells. 
 A proliferating cell in compartment $i$ divides and the two daughter cells differentiate and migrate into the next downstream compartment ($i+1$) with probability $\varepsilon$, increasing the downstream compartment by 2 cells, mutates with probability $u$ or self renews within its own compartment with probability $1-\varepsilon-u$. Mutated cells stay in the hierarchy. 
 If a mutated cell proliferates, it 
 differentiates 
 with probability $\varepsilon$ into the next downstream compartment, 
 it
 self renews with probability $1-\varepsilon-u$, or it mutates with probability $u$ again, leading to a cell with two (or more) mutations. 
 All possible outcomes of a cell proliferation are depicted in figure \ref{Network}. The direction of the arrows point towards the accessible cell states and the labels give the transition probabilities.
 We allow arbitrary parameters and introduce $\varepsilon_{i}^{k}$ as differentiation probability of cells in compartment $i$ carrying $k$ mutations. Asymmetric cell divisions are not explicitly implemented, as they can be absorbed in the differentiation probabilities on the population level. 
  The fate of a cell's offspring is determined based on the probabilities $\varepsilon_{i}^{k}$.
 Cells proliferate with a rate $r_{i}$ in each compartment $i$. 
 Usually cells in upstream compartments proliferate slow and cell proliferation speeds up in downstream compartments (i.e. $r_{i} < r_{i+1}$). 
 This general framework is very flexible and different tissue structures can be represented.

 \subsection{Stochastic individual based simulations}
We implement individual based stochastic simulations of the cell dynamics in hierarchically organized tissue structures. We utilize an implementation of the Gillespie algorithm \citep{gillespie:JoCP:1976,gillespie:ARPC:2007}. Originally introduced to simulate chemical reactions, it allows us to reproduce exact stochastic trajectories of the system. Each cell has an individual representation. Thus, the complete clonal history of cells within the hierarchy can be recorded. If a cell is chosen for reproduction (determined by the Gillespie method), it either differentiates, self renews, or mutates according to the probabilities $\varepsilon_{i}^{k}$ and $u$. The parameters of the simulated system are described below and chosen to represent human hematopoiesis.

\subsection{The hematopoietic system}
 
 In the following we focus on the hematopoietic system.  There, about $400$ stem cells replenish the hematopoietic cell pool \citep{dingli:PlosOne:2006,buescher:JCI:1985}. Each stem cell divides approximately once a year \citep{rufer:JEM:1999,dingli:PlosOne:2006}. Cell proliferation is assumed to increase exponentially $r_{i}=\gamma^{i}r_{0}$, with $\gamma=1.26$ and $r_{0}$ corresponds to the proliferation rate of stem cells. The differentiation probability is assumed to be constant,  $\varepsilon=0.85$, for all non stem cell compartments and in total $i=31$ compartments are needed, to ensure a daily bone marrow output of approximately $3.5\times 10^{11}$ cells \citep{dingli:PlosOne:2007,werner:PlosCB:2011}.

  \section{Results}

\subsection{Time continuous dynamics of multiple mutations}
 
\begin{figure}
\center
    \includegraphics[width=0.6\textwidth]{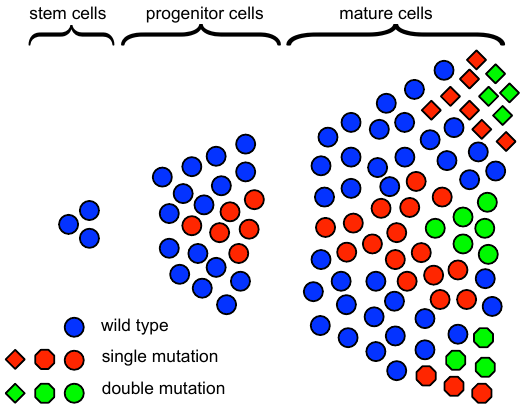}
\caption{Clonal expansion within a hierarchically organized tissue. Cell proliferation is driven by a few slow dividing stem cells, giving raise to faster dividing progenitor cells. After some differentiation steps the mature tissue cells are obtained. Initially cells have
no mutations, but mutants can arise and expand within the hierarchy. These cells either vanish or gain an additional mutation, which again potentially spreads within the hierarchy.
Different colors code for a different number of mutations, whereas different shapes indicate different mutations. 
}
\label{Nz2}
\end{figure}

  We describe the deterministic dynamics of a cell population within a hierarchically organized tissue structure, that initially carries no mutation. 
 A cell may commit further into the hierarchy (differentiate), mutate, or self renew itself. 
  This occurs with probability $\varepsilon$, $u$ and $1-\varepsilon-u$, respectively. In figure \ref{Network} a schematic representation of the resulting hierarchical structure is shown. Compartments to the right represent downstream compartments of more specialized (differentiated) cells, while compartments to the bottom represent states of cells, that accumulated an additional mutation. During one cell division, a cell either mutates and moves one compartment to the bottom, differentiates and produces two cells in the next downstream compartment to the right, or self renews and produces an additional cell within its original compartment. This leads to an expansion of clonal populations within the hierarchy, that potentially accumulates several (distinct) mutations during the differentiation process. This is schematically shown in figure \ref{Nz2}.
  
 The above transition probabilities can be used in an individual based stochastic simulation. In the following, we provide a deterministic description of the average dynamics of cells carrying multiple mutations in such hierarchical structures. 
 Thus, we describe the dynamics by transition rates instead of transition probabilities,
 but only averages enter in our description. By doing so, we neglect certain effects, such as stochastic extinction of cells. However, the approach allows us to investigate the averages of the underlying stochastic simulations.
  
 Assume that in compartment $i$ there are $N_{i}^{k}\left(t\right)$ cells carrying $k$ mutations at time $t$. The number of these cells increases due to influx from the upstream compartment at a rate $2r_{i-1}\varepsilon_{i-1}^{k}$, mutations with a rate $r_{i}u$ and self renewal at a rate $r_{i}\left(1-\varepsilon_{i}^{k}-u\right)$. Cells are lost either by mutation at a rate $r_{i}u$, or differentiation at a rate $r_{i}\varepsilon_{i}^{k}$. The deterministic description of the hierarchical compartment model becomes a system of coupled differential equations \citep{werner:PlosCB:2011}, given by
 \begin{equation}
\dot N_{i}^{k}\left(t\right)=\begin{cases}
r_{i}\left(1-2\alpha_{i}^{k}\right)N_{i}^{k}\left(t\right)+2r_{i}\varepsilon_{i-1}^{k} N_{i-1}^{k}\left(t\right) & k=0 \\
r_{i}\left(1-2\alpha_{i}^{k}\right)N_{i}^{k}\left(t\right)+2r_{i}\varepsilon_{i-1}^{k} N_{i-1}^{k}\left(t\right)+r_{i}uN_{i}^{k-1}\left(t\right) & k>0.\end{cases} \label{DifEqu}
\end{equation}
Here $\alpha_{i}^{k}=\varepsilon_{i}^{k}+u$ denotes the probability that a cell with $k$ mutations leaves compartment $i$. 
Typically, $\alpha_i^k$ is very close to $\varepsilon$.
A model for stochastic cell dynamics in the stem cell compartment for neutral and non neutral mutations can be found in \citep{dingli:CCy:2007,traulsen:JRSI:2013}. In that paper, the stochastic Moran process is used to investigate the extinction and fixation probabilities of stem cell mutations. The deterministic stem cell driven cell replenishment in hierarchical tissues is studied in detail in 
\cite{werner:PlosCB:2011}.
However, in that prior work, the effects arising from additional mutations were neglected. 
Here we focus on non stem cell driven clonal dynamics. We explicitly allow for an arbitrary number of mutational hits at any stage of the hierarchy, but we neglect a continuous influx of mutated cells from the stem cell level.
This assumption gives the condition $N_{0}^{k}\left(t\right)=0$. The initial condition
\begin{equation}
N_{i}^{k}\left(0\right)=\begin{cases} n_{0} & i=1\; \;\;k=0 \\
0 & \mathrm{otherwise},\end{cases}
\end{equation}
corresponds to initially $n_{0}$ cells in compartment 1 carrying no mutation. One can think of a neutral marker approach, where one cell in the hierarchy is genetically marked, and one considers the clonal population arising from this marked cell \citep{gerrits:Blood:2010}. 
Although we neglect a continuous influx of mutated cells from the stem cell compartment, stem cell mutations can be implemented indirectly. Our approach allows for altered cell proliferation properties of the founder cell, potentially derived by a mutation at the stem cell level. 
 For a constant differentiation probability, 
 i.e.\ the case where all $\alpha_i^k$ and all $\varepsilon_i^k$ are identical,
 equation \eqref{DifEqu} can be solved recursively.  
The number of cells in compartment $i$ carrying no mutation changes in time as
\begin{equation}
N_{i}^{0}\left(t\right)=n_{0} \frac{\left(2\varepsilon\right)^{i-1}}{\left(2\alpha-1\right)^{i-1}}\left(\prod_{j=1}^{i-1}r_{j}\right)\sum_{l=1}^{i}\frac{\mathrm{e}^{r_{l}\left(1-2\alpha\right)t}}{\prod\limits_{\substack{h=1 \\ h\neq l}}^i\left(r_{h}-r_{l}\right)}. \label{sol1}
\end{equation} 
The solution can also be derived recursively for cells carrying $k$ mutations in compartment $i$ and becomes 
\begin{align}
N_{i}^{k}\left(t\right)=n_{0}\sum_{h=0}^{k}\frac{u^{k-h}g_{i}^{h}}{k!}\frac{\left(2\varepsilon\right)^{i-1}}{\left(2\alpha-1\right)^{i+h-1}}\left(\prod_{j=1}^{i-1}r_{j}\right)\sum_{l=1}^{i}\frac{\left(r_{l}\,t\right)^{k-h}\mathrm{e}^{r_{l}\left(1-2\alpha\right)t}}{\prod\limits_{\substack{h=1 \\ h\neq l}}^i\left(r_{h}-r_{l}\right)}, \label{solution}
\end{align}
where $g_{i}^{h}$ is a combinatoric parameter denoted in \ref{comtab} for cells carrying up to three mutations,
 \begin{equation}
 \begin{array}{|c|c|c|c|c|}\hline
 & h=0 &Êh=1 & h=2 & h=3  \\  \hline
ÊÊÊÊk=0ÊÊÊÊ & 1 &  &  & \\ 
ÊÊÊÊÊÊk=1ÊÊ & 1 & i-1 &  & \\
Êk=2ÊÊ & 1 & (i-1)k & i(i-1) & \\
Êk=3ÊÊ & 1 & (i-1)k& i(i-1)k & (i+1)i(i-1)  \\
\hline
 \end{array}
 \label{comtab}
 \end{equation}
If $\alpha<0.5$, non stem cells will continuously accumulate in downstream compartments.
The probability of self renewal in this case is larger than the probability of differentiation. This scenario seems to be realized in certain blood cancers. For example in acute promyelocytic leukemia an abnormal increase of immature granulocytes and promyelocytes is observed, resulting from a block of cell differentiation at a late progenitor cell stage \citep{raymond:NEJM:1993,guibal:Blood:2009}. However these cases are rare.      

\begin{figure}
\center
   \includegraphics[width=0.99\textwidth]{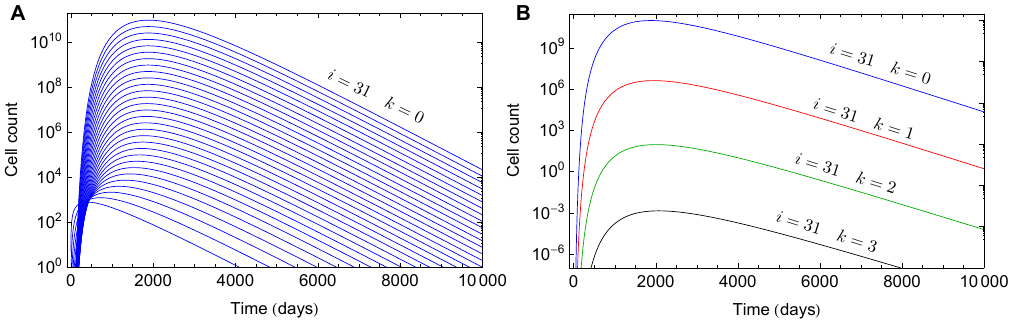}
\caption{\textbf{A} Number of cells carrying no mutation in compartments 1 to 31 arising from compartment 1 containing $1000$ cells. Lines show equation \eqref{sol1}, with parameters $n_{0}=1000$, $\varepsilon=0.85$, $\gamma=1.26$, $u=10^{-6}$ and $r_{0}=1/400$. Cells are more likely to differentiate than to self renew and thus progressively travel into more committed compartments. Initially, the cell count increases, but cells get washed out in the long run. The time scale is determined by the number of stem cell divisions. A stem cell is assumed to divide once a year, thus after 400 stem cell divisions a year passed. \textbf{B} Count of cells carrying zero to three mutations in compartment 31, given by equation \eqref{solution}. We used same parameters as in A. Cells carrying multiple mutations are exponentially suppressed. }
\label{Mutant}
\end{figure}

For $\alpha>0.5$, the solution becomes a clonal wave, traveling through the hierarchy in time.
In this case the probability of differentiation is larger than the probability of self renewal and thus cells progressively travel downstream, see figure \ref{Mutant} as an example. The cell population founded by a single non stem cell expands within the hierarchy initially, but gets washed out and vanishes in the long run. This is believed to be true for healthy homeostasis. For example for the hematopoietic system the differentiation probability was estimated to be $\varepsilon=0.85$ \citep{dingli:PlosOne:2007}. As by far most cell proliferations occur at the progenitor and more committed differentiation stages, this provides a natural protection of the organism against the accumulation of multiple mutations, as the survival time of most (non stem cell like) mutations is finite.

 The maximum mutant cell count of the clonal wave and the time to reach this maximum can be calculated for the compartment of the mutant origin, in our case the first compartment. The time is given by 
 \begin{equation}
 t_{\mathrm{max}}^{k}=\frac{k}{\left(2\alpha-1\right)r_{1}}.
 \end{equation} 
  The time to reach the maximum increases linearly with the number of additional mutations $k$. 
The cell count at the maximum becomes
 \begin{equation}
 N_{1}^{k}\left(t_{\mathrm{max}}\right)=\frac{\left(uk\right)^{k}}{k!\left(2\alpha-1\right)^{k}}\mathrm{e}^{-k}\approx \frac{u^{k}}{\left(2\alpha-1\right)^{k}}\frac{1}{\sqrt{2\pi k}},
 \end{equation}
where we used the Stirling formula to approximate 
k!.
The maximum scales with $u^{k}$ and
 thus decreases exponentially with $k$.
  In addition the factor $1/\sqrt{k}$ leads to a further suppression of the maximum for increasing $k$. However, the risk of additional mutations depends not on the maximal cell count, but on the reproductive capacity of a cell line. This reproductive capacity can be captured by the cumulative cell count. The number of cells within compartment $i$ carrying $k$ mutations produced until time $t$ is given by  
 \begin{equation}
 m_{i}^{k}\left(t\right)= r_{i}\alpha_{i}^{k}\int_{0}^{t}\mathrm{d}s\, N_{i}^{k}\left(s\right), \label{Integral}
 \end{equation}
 and the reproductive capacity can be derived by taking the time limit to infinity. The general solution \eqref{solution} allows to carry out the integral exactly by integration by parts. However, the problem can be tackled from a different perspective, leading to a more transparent solution of \eqref{Integral} that is easier to handle.

\subsection{Cell reproductive capacity}
 
We call the cell subpopulation within a compartment $i$, that is derived by a single founder cell in an upstream compartment, the reproductive capacity of this founder cell. This idea directly corresponds to the method of neutral markers. We imagine a genetically marked cell somewhere in the hierarchy and count the offspring of this cell at any stage of the hierarchy. This corresponds to the total count of cells with same color in figure \ref{Nz2}. 

Assume a single cell carrying no mutation in compartment 1. This cell differentiates with probability $\varepsilon_{1}^{0}$ into the next downstream compartment, mutates with probability $u$ or produces an additional cell in compartment 1 with probability $1-\varepsilon_{1}^{0}-u$. We first discuss the probability, that a cell leaves compartment 1 after exactly $l$ cell divisions. A cell can leave a compartment either by mutation or differentiation, before which the cell has to undergo $l-1$ self renewals. 
Thus, this probability becomes $\left(\varepsilon+u\right)\left(1-\varepsilon-u\right)^{l-1}$. 
During this time, the cell population in compartment 1 derived from this single cell increased to $2^{l-1}$ cells, if all daughter cells share the same proliferation probabilities. With this, the reproductive capacity of a single cell in compartment 1 is  on average
 \begin{equation}
m_{1}^{0}=\sum_{l=0}^{\infty}\alpha_{1}^{0}\:2^{l}\left(1-\alpha_{1}^{0}\right)^{l}=\begin{cases}\frac{\alpha_{1}^{0}}{2\alpha_{1}^{0}-1} & \alpha_{1}^{0} > \frac{1}{2} \\
\infty & \alpha_{1}^{0} \le \frac{1}{2}.\end{cases} \label{01}
\end{equation}
The sum becomes infinite if $\alpha \le 0.5$, as the probability to produce offspring in the founder compartment is higher than the probability to leave the compartment. 
Thus, the cell population continuously increases. 
Of course, under normal conditions, cells do not have an unlimited capacity to divide and serial telomere erosion amongst others will impose a physical limit on the number of divisions a cell can undergo \citep{hayflick:ECR:1961, armanios:NRG:2012}. 
The biologically more relevant case is $\alpha>0.5$ and cells tend to differentiate into more committed compartments. 
In this case, the total number of offspring cells that arise from a single cell (i.e. a clone) is finite and given by \eqref{01}. 
The number of cells $m_{i}^{0}$ in compartment $i$ carrying no mutation, increases due to influx of cells via differentiation from compartment $i-1$ and the expansion of these cells due to self renewal in compartment $i$. Thus, we can write for the reproductive capacity of cells in compartment $i$ without mutation
\begin{equation}
m_{i}^{0}=2\frac{\varepsilon_{i-1}^{0}}{\alpha_{i-1}^{0}}m_{i-1}^{0}\frac{\alpha_{i}^{0}}{2\alpha_{i}^{0}-1}=\frac{\alpha_{i}^{0}}{2\alpha_{i}^{0}-1}\prod\limits_{l=1}^{i-1}\frac{2\varepsilon_{l}^{0}}{2\alpha_{l}^{0}-1}. \label{healthy}
\end{equation}
This can be generalized, and an expression for the reproductive capacity of cells in compartment $i$ carrying $k$ mutations can be derived. Cells in compartment $i$ with $k$ mutations are acquired either by differentiation of cells from compartment $i-1$ that carry $k$ mutations, by mutation of cells in compartment $i$ carrying $k-1$ mutations, or by self-renewal of cells already in compartment $i$ and $k$ mutations. 
With this, we can write 
\begin{equation}
m_{i}^{k}=\left(2\frac{\varepsilon_{i-1}^{k}}{\alpha_{i-1}^{k}}m_{i-1}^{k}+\frac{u}{\alpha_{i}^{k-1}}m_{i}^{k-1}\right)\frac{\alpha_{i}^{k}}{2\alpha_{i}^{k}-1},
\end{equation}
 where the two terms in the bracket represent cells either produced by differentiation or mutation, multiplied with the self renewal potential of these cells. This recurrence relation can be solved recursively
 \begin{equation}
 m_{i}^{k}  =\frac{\alpha_{i}^{k}}{2\alpha_{i}^{k}-1}\sum_{l=1}^{i} 2^{i-l} \frac{u}{\alpha_{l}^{k-1}}m_{l}^{k-1}\prod\limits_{h=l}^{i-1}\frac{\varepsilon_{h}^{k}}{2\alpha_{h}^{k}-1}. \label{general}
 \end{equation}
 Since $m_{i}^{0}$ is given by \eqref{healthy} one can construct the explicit solution iteratively. Equation \eqref{general} allows for arbitrary parameters $\alpha_{i}^{k}$ and thus incorporates any mutational induced change in cell proliferation parameters. However, if $\alpha_{i}^{k}\le 0.5$ the sum diverges. Cells with at least $k$ mutations will accumulate in all compartments downstream $i$. 
 Note, that equation \eqref{general} captures the general deterministic dynamics of a cell lineage founded by a single cell somewhere in the hierarchy. 
 We are especially interested in the case where this founder cell carries critical, potentially cancer driving mutations that will allow us to address the expected number of mutations arising in this mutant clone. 
 In addition, also the probability of obtaining such a critical mutational hit can be investigated. 
 Not that equation \eqref{general} is based on a fully deterministic picture. 
But even if a cell accumulated a critical number of mutations, it might still go extinct du to stochastic effects, see for example \citep{beerenwinkel:PlosCB:2007,lenaerts:HAE:2010,traulsen:JRSI:2013}.
In the following, we discuss the general solution of equation \eqref{general} for mutations that are neutral relative to the founder cell.  
  
\subsection{Reproductive capacity of neutral mutants} 

\begin{figure}
  \center
    \includegraphics[width=0.99\textwidth]{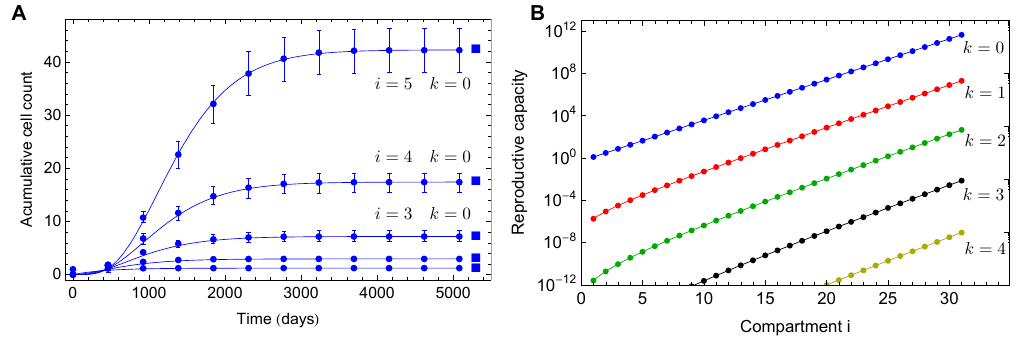}
\caption{\textbf{A} Number of cells without mutations in compartments 1 to 5, arising from a single cell in compartment 1. Lines are equation \eqref{Integral}, symbols are averages with corresponding standard deviations over $10^{3}$ independent runs of stochastic individual based computer simulations and squares are equation \eqref{Neutral}. Parameters are $n_{0}=1$, $\varepsilon=0.85$, $\gamma=1.26$, $u=10^{-6}$ and $r_{0}=1/400$. \textbf{B} Reproductive capacity of a single founder cell in compartment 1. Shown is the number of cells with 0 to 4 mutations in the first 31 compartments, acquired from a single cell in compartment 1. Symbols are numerical solutions of \eqref{Integral} in the limit of infinite time and lines are equation \eqref{Neutral}. The reproductive capacity increases exponentially for increasing compartment number. Cells carrying multiple mutations are strongly suppressed within a hierarchical tissue structure, see equation \eqref{Ratio}.}
\label{AcCell}
\end{figure}

We call a mutation neutral if the reproductive capacity of the mutant and the founder cell is equal. 
In the former subsection we have shown, that the reproductive capacity of a cell depends on its differentiation probability $\varepsilon$ and its mutation rate $u$, but interestingly it is independent of the reproduction rate $r$. 
Therefore the clonal lineage and the number and type of mutations that arise from a single founder cells do not depend on the proliferation rates of the founder cells. 
However, the time to reach those states of course depends on $r$. 
Therefore, although two mutations lead to the same outcome, this might occur on distinct time scales, with observable differences in the progression of diseases. 
Nonetheless, our definition of neutral mutations only requires constant differentiation probabilities and mutation rates relative to the founder cell.  
This assumption allows us to write $\varepsilon_{i}^{k}=\varepsilon_{i}$ and thus the number of parameters is reduced from $\left(k+1\right)i+1$ for the general case to $i+1$ for the neutral case. This number can be reduced to two parameters $u$ and $\varepsilon$, if a constant differentiation probability for all non stem cell stages is assumed, $\varepsilon_{i}=\varepsilon$. This simplifies the evaluation of the recurrence relation \eqref{general} significantly. The reproductive capacity $m_{i}^{k}$ of neutral mutations in compartment $i$ carrying $k$ mutations becomes 
\begin{equation}
m_{i}^{k}=\alpha\frac{u^{k}}{k!}
\frac{\left(2\varepsilon\right)^{i-1}}{\left(2\alpha-1\right)^{i+k}}\prod_{l=1}^{k}\left(i+l-1\right)=\alpha u^{k}\frac{\left(2\varepsilon\right)^{i-1}}{\left(2\alpha-1\right)^{i+k}}{i+k-1\choose k}. \label{Neutral}
\end{equation}
 Mutants carrying $k$ mutations are suppressed by a factor $u^{k}$ and thus are rare in early differentiation stages. The number increases exponentially for downstream compartments, and a significant load of cells carrying few mutations can be observed in late differentiation stages, see figure \ref{AcCell}. 
 
Equation \eqref{Neutral} reveals interesting properties of hierarchical tissue structures. The ratio of cells carrying $k$ mutations to cells carrying $k-1$ mutations in compartment $i$ is 
 \begin{equation}
\frac{m_{i}^{k}}{m_{i}^{k-1}}=\frac{u}{2\alpha-1}\left(1+\frac{i-1}{k}\right) . \label{Ratio}
\end{equation}
 The ratio increases with compartment number, but the increase becomes flatter for increasing $k$.  The compartment structure leads to an additional suppression of cells carrying multiple mutations and thus is a protection mechanism against cancer initiation. 
 The ratio is constant for $i=1$, the compartment of the mutant origin. The protection mechanism affects downstream compartments only. 
 
 On the other hand, the scaling properties for more differentiated cells show interesting properties also. The ratio of cells with $k$ mutations in compartment $i+1$ to cells with $k$ mutations in compartment $i$ is given by
\begin{equation}
\frac{m_{i+1}^{k}}{m_{i}^{k}}=\frac{2\varepsilon}{2\alpha-1}\left(1+\frac{k}{i}\right).
\end{equation}  
 The increase of cells is constant for cells carrying no mutations $k=0$. It increases with $k$, but is suppressed within the hierarchy by the factor of $1/i$. 
\subsection{Number of distinct neutral mutations}
 
So far, we have discussed the reproductive capacity of cells. However, we did not distinguish between different mutations, but grouped together cells with an equal number of mutations. Often experimental studies focus on mutation landscapes, investigating the variation in clonal loads in healthy and sick individuals. 
Our approach allows us to estimate the expected number of distinct mutations that arise from a single founder cell, corresponding to the count of distinct symbols with same color in figure \ref{Nz2}.  
 
Let us assume that every mutation event is unique. Thus, we neglect the possibility that the same mutation is derived twice independently. The diversity in compartment $i$ increases either by additional mutations of cells in compartment $i$, or differentiation of clones from compartment $i-1$ into compartment $i$. Assuming a differentiation probability $\varepsilon>0$ the expected diversity $n^{k+1}_{i}$ of cells with $k+1$ mutations in compartment $i$ is 
\begin{equation}
n_{i}^{k+1}=u\sum_{j=1}^{i}m_{j}^{k}= \alpha u^{k+1}\sum_{j=1}^{i}\frac{\left(2\varepsilon\right)^{j-1}}{\left(2\alpha-1\right)^{j+k}}{j+k-1\choose k}.\label{diver}
\end{equation}   
As an example, if we use the parametrization of the hematopoietic system, $\varepsilon=0.85$ and $u=10^{-6}$, we find approximately  $n_{31}^{1}\approx 30$ distinct single mutations in compartment 31,
 that were derived from the clonal progeny of a single founder cell in compartment 1. But the founder cell by chance could carry a mutation, that changes their cell proliferation parameters. For example, the differentiation probability of the founder cell could change to $\varepsilon=0.75$. In this case, equation \ref{diver}  
 estimates 28000 distinct mutations derived from this single founder cell. We note, that the above change of the differentiation probability 
 from 0.85 to 0.75
 is sufficient to explain the manifestation of chronic myeloid leukemia in otherwise healthy adults \citep{dingli:CLEU:2008,Lenaerts:CCY:2011}.
 However, equation \eqref{diver} represents an average and in individual cases fluctuations, caused by stochasticity are expected. But also small changes in $\varepsilon$ or $u$ influence the expected diversity significantly. A linearized error analysis \citep{taylor:book} reveals the dependency of \eqref{diver} on the uncertainties of $u$ and $\varepsilon$, which is given by
\begin{equation}
\Delta n_{i}^{k+1}=\sum_{j=1}^{i}\frac{\left(2\varepsilon\right)^{j-1}}{\left(2\alpha-1\right)^{j+k+1}}\left(g_{j}^{\varepsilon}+g_{j}^{u}\right){j+k-1\choose k}, \label{error}
\end{equation}
 with $g_{j}^{\varepsilon}=u^{k+1}|2\alpha\left(1-j-k\right)-1|\Delta \varepsilon$ and 
 \newline $g_{j}^{u}=u^{k}|(k+1)\varepsilon\left(2\varepsilon-1\right)+u\left(2\left(k-j+3)\right)\varepsilon-k-2\right)+2u^{2}\left(2-j\right)|\Delta u $. 
 If we assume $\Delta u= 10^{-7}$ and $\Delta \varepsilon=0.01$ the uncertainty given by \eqref{error} becomes $\Delta n_{31}^{1}\approx35$, where the individual error contributions in $u$ and $\varepsilon$ are $7$ and $28$ respectively. Especially note the strong dependency on $\Delta \varepsilon$. If we choose $\Delta \varepsilon=0.05$, a deviation, that might be difficult to detect in vivo, one gets $\Delta n_{31}^{1}\approx 150$. Thus, small variations in $\varepsilon$ lead to significant differences in the expected diversity of clonal populations, one aspect that might contribute to the explanation of the observation of very diverse mutation landscapes. Note also, that an increasing mutation rate $\Delta u=10^{-6}$ gives  $\Delta n_{31}^{1}\approx 65$. Of course, higher mutation rates increase the expected diversity of clonal populations. However, a higher mutation rate (or genomic instability) is neither the exclusive nor necessarily the dominant underlying cause of the diversity in the mutation landscape that is observed.

\subsection{Example: clonal diversity in acute lymphoblastic leukemia}
 
Let us now consider a specific example using data from childhood acute lymphoblastic leukemia (ALL). The most common chromosomal abnormality in this disease is the t(12;21) translocation that results in the fusion gene ETV6-RUNX1 (also known as \\TEL/AML1). There is evidence that this mutation often arises in utero. This has been confirmed to be the case in at least one pair of monozygotic twins \citep{hong:Science:2008}. This mutation is a founder mutation and is considered to be critical for the disease. Cells that express this fusion gene appear to have a higher self-renewal and enhanced survival compared to normal cells \citep{fischer:Oncogene:2005,hong:Science:2008}. In our model,  enhanced self-renewal implies a reduced differentiation probability for the cells carrying the mutation 
($\varepsilon<0.85$).
Recently, Ma et al. performed whole genome sequencing on leukemic cells isolated from two pairs of monozygotic twins. In one pair of twins, the initial event occurred in utero since the ETV6-RUNX1 fusion was shared by both siblings \citep{ma:PNAS:2013}. They found that the incidence of non-synonymous single nucleotide mutations between the samples ranged from 708 to 1237 \citep{ma:PNAS:2013}. These mutations must have occurred after the founder mutation and independent of each other. The time from the putative appearance of the shared ETV6-RUNX1 mutation and disease was 48 to 55 months.  
The second pair of monozygotic twins shared a mutation in NF-1, and the time to diagnosis of ALL was 72 to 77 months after birth. The tumors in these two children had 949 to 975 unique non synonymous single nucleotide mutations.
Using these constraints
and an estimate of 10\%-75\% cancer cells at diagnosis, unchanged proliferation rates  as well as $N_0=100$ \cite{dingli:PRSB:2007}, 
we use equation \eqref{solution} to estimate the differentiation probability $\varepsilon$ for the mutant cells, which is then in the range of $0.78$ to $0.81$, ie.\ only slightly lower than that of normal cells.  
Based on equation \eqref{solution}, it takes approximately 50 to 80 months to reach this load. 
We further predict a range of 350 to 2600 distinct mutations by utilizing equation \eqref{diver}, assuming a mutation rate of $u=10^{-6}$ and the above range of differentiation probabilities. 
Thus, we expect slightly more distinct mutations than found in patients. 
But some of those theoretically predicted clonal populations are small and may escape detection. 
For example, if we neglect mutations, that occur during the last differentiation step of cells in compartment 30 to 31 (expected to be small cell populations), we predict only 150 to 950 distinct mutations.
Clearly, the model as presented can explain the large number of passenger mutations that can be expected in a typical patient with ALL and likely other types of leukemia.

 \section{Discussion}

The accumulation of multiple mutations in cells is considered to be critical for cancer initiation. 
However, mutations unavoidably accompany cell proliferation and thus cancer can potentially occur in any multicellular organism. 
Hierarchical tissue structures contribute to the protection against such mutations.
So far, the suppression of single mutations in hierarchical tissue structures has been the focus. Here, we have shown that in addition, the risk of the accumulation of multiple mutations is dramatically reduced by a hierarchical tissue organization. The cell population is divided into few slow proliferating stem cells and many faster proliferating progenitor cells. Stem cells have an almost infinite cell reproductive capacity, but the manifestation of a critical mutational load often exceeds an organisms expected natural life time. 
Progenitor cells proliferate faster, but their reproductive capacity usually is limited and thus they give rise to clonal waves traveling through the hierarchy. Still both cases can be observed. There are cancers that presumably originate from stem cell mutations for example chronic myeloid leukemia, and there are cancers that originate in later stages of hematopoiesis, for example acute promyelocytic leukemia and various other subtypes of acute myeloid leukemia \citep{guibal:Blood:2009}. Understanding the clonal dynamics for both cases is of importance. 
Here, we have focused on the accumulation of non stem cell driven mutations in hierarchically organized tissues. We arrived at closed solutions for the clonal waves traveling through the hierarchy. From this the reproductive capacity of cells can be deduced. This allows us to predict the expected risk to acquire any number of mutations from one single cell, given the proliferation properties of this cell. We derived equations that allow the quantitative classification of multiple mutations in hierarchically organized structures, highlighting the strong suppression of clones carrying multiple mutations by this architecture. 
Although we neglected clonal competition over limited resources, such as cytokines or nutrients, we expect the model to capture general patterns of clonal expansions in hierarchically organized tissues and serving as a null model. 
Moreover, those cancer cells that divide independently of proliferation signals (for example due to mutated tyrosine kinases) escape such a competition.

Another important question emerged more recently with the accessibility of whole genome sequencing technology. These techniques revealed very complex and diverse mutation landscapes for many different cancers, with the classification of driver and passenger mutations as the final goal. Knowing the exact driver mutations might help to understand the properties of specific cancer cells, allowing the development of effective treatment strategies. A promising example is the design of molecularly targeted agents such as the various tyrosine kinase inhibitors (Imatinib, Nilotinib etc) to treat patients with chronic myeloid leukemia. These molecules specifically bind to kinase domains, encoded by the BCR-ABL oncogene and strongly suppress the proliferative capacity of these cells \citep{druker:NEJM:2006,saglio:NEJM:2010,werner:PLosOne:2011}.
Our work can contribute to this question by predicting the average number of distinct neutral (passenger) mutations, acquired from a single cell at any stage of the hierarchy. This approach directly corresponds to the method of a neutral marker. The genetically marked cell represents the founder cell of the clonal population and one follows the offspring of this founder cell throughout the hierarchy. This enables the prediction of the size and the variability of the clonal population. 
For example, in normal hematopoiesis, we expect cells to have a differentiation probability of $\varepsilon=0.85$, leading to approximately 30 distinct single mutations (subclones) in adult cells, acquired from this single cell.
If this founder cell acquired a mutation, that changed the differentiation probability to $\varepsilon=0.75$ by chance, the expected number of distinct single mutations increases to approximately 28000.

We have shown how even a slight change in the self-renewal probability of progenitor cells can lead to substantial differences in the number of passenger mutations observed in ALL. This likely holds true for other malignancies.  In acute promyelocytic leukemia, apart from the t(15;17) that is a critical event in the origin of this disease (akin to the ETV6-RUNX1 discussed above for ALL), approximately 440 non synonymous single nucleotide mutations were found that were unique to the tumor clone \citep{welch:Cell:2012}. Interestingly, it is highly likely that the cell of origin of APL is downstream (a progenitor cell) of the cell of origin of ETV6-RUNX1 driven ALL and this may, in part, explain the less diverse mutational landscape reported in APL compared to ALL and would fit well with our model. Of course, any genomic instability will further increase the repertoire of passenger mutations that is observed in any given tumor.

We also note, that the effect of a specific mutation on a cell needs not be large for the effect to spread throughout the tumor. Tissue architecture and dynamics as in hematopoiesis serve as a deterrent against the accumulation of mutations, 
in particular multiple mutations occurring in one cell.
Once a driver mutation appears, if this changes either the self-renewal of the cell or the mutation rate, then the appearance of many passenger mutations becomes inevitable in such an architecture due to the amplification of cells that occurs. Thus, a minor change in the differentiation probabilities that might be difficult to detect in vivo drastically changes the expected number of passenger mutations that would be observed. Concomitantly, this will increase the risk of acquiring mutations in 'driver genes' and so lead to malignancy.
However, if the initiation of disease requires several co-occurring mutations, a hierarchical tissue structure is a powerful mechanism of tumor suppression.

\acknowledgements{B.W.\ and A.T.\ thank the Max Planck Society and the Emmy-Noether program of the German Research Foundation for generous funding. }

\end{document}